\documentclass[trackchanges]{aastex701}

\newcommand{\degree}{^{\circ}}
\newcommand{\tO}{$T_0$}
\newcommand{\te}{$T_e$}
\newcommand{\tm}{$T_m$}

\newcommand{\Fig}[1]{Figure~\ref{#1}}

\def\blue{\textcolor{blue}}

\begin{document}

\title{Observed Joy's law of Bipolar Magnetic Region tilts at the emergence supports the thin flux tube model}

\author[orcid=0000-0001-7036-2902,sname='Sreedevi']{Anu Sreedevi}
\affiliation{Department of Physics, Indian Institute of Technology (Banaras Hindu University), Varanasi, 221005, India}
\email[show]{anubsreedevi.rs.phy20@itbhu.ac.in}  

\author[orcid=0000-0002-8883-3562, sname='Karak']{Bidya Binay Karak} 
\affiliation{Department of Physics, Indian Institute of Technology (Banaras Hindu University), Varanasi, 221005, India}
\email[show]{karak.phy@iitbhu.ac.in}

\author[orcid=0000-0003-3191-4625, sname='Jha']{Bibhuti Kumar Jha}
\affiliation{Southwest Research Institute, Boulder, CO 80302, USA}
\email{bibhuti.jha@swri.org}
\author[sname='Gupta']{Rambahadur Gupta}
\affiliation{Department of Physics, Indian Institute of Technology (Banaras Hindu University), Varanasi, 221005, India}
\email{rambahadurgupta.rs.phy23@itbhu.ac.in}

\author[orcid=0000-0003-4653-6823,sname='Banerjee']{Dipankar Banerjee}
\affiliation{Department of Earth and Space Sciences, Indian Institute of Space Science \& Technology, Thiruvananthapuram 695547, Kerala, India}
\affiliation{Indian Institute of Astrophysics, Koramangala, Bangalore 560034, India}
\affiliation{Center of Excellence in Space Sciences India, IISER Kolkata, Mohanpur 741246, West Bengal, India}
\email{dipu@iist.ac.in}

\begin{abstract}

Bipolar sunspots, or more generally, Bipolar Magnetic Regions (BMRs), are the dynamic magnetic regions that appear on the solar surface and are central to solar activity. One striking feature of these regions is that they are often tilted with respect to the equator, and this tilt increases with the latitude of appearance, popularly known as Joy's law. Although this law has been examined for over a century through various observations, its physical origin is still not established. 
An attractive theory that has been put forward behind Joy's law is the Coriolis force acting on the rising flux tube in the convection zone, which has been studied using the thin flux tube model. 
However, observational support for this theory is limited. 
If the Coriolis force is the cause of the tilt, then we expect BMRs to hold Joy's law at their initial emergence on the surface. By automatically identifying the BMRs over the last two solar cycles from high-resolution magnetic observations, we robustly capture their initial emergence signatures on the surface. We find that from their appearance, BMRs exhibit tilts consistent with Joy’s law. This early tilt signature of BMRs suggests that the tilt is developed underneath the photosphere, driven by the Coriolis force and helical convection, as predicted by the thin flux tube model.
Considerable scatter around Joy's law observed during the emergence phase, which reduces in the post-emergence phase, reflects the interaction of the vigorous turbulent convection with the rising flux tubes in the near-surface layer. 
\end{abstract}

\keywords{\uat{Bipolar sunspot groups}{156} --- \uat{Solar activity}{1475} --- \uat{Solar magnetic fields}{1503} --- \uat{Solar active region magnetic fields}{1975} --- \uat{Solar physics}{1476}}

\section{Introduction} 
One of the most striking features of the Sun’s visible surface is the appearance of dark spots, famously known as sunspots. 
Advances in solar observations revealed that sunspots host intense magnetic fields, with adjacent spots possessing opposite polarities \citep{H08}. This led to the broader term, Bipolar Magnetic Regions (BMRs), to describe them. Given their central role in driving solar activity, understanding the variabilities in solar cycles has long been a focus of research \citep{K2023}.  Although years of studies show that surface BMRs are expressions of a smoothly operating solar dynamo underneath \citep{Kar14a, Cha20}, their specific characteristics, such as emergence and orientation, remain active areas of investigation and debate \citep{F2009, LG2015}. 

A distinctive property of the emerging BMRs is that the line connecting their opposite magnetic polarities forms an apparent tilt angle with respect to the equator. Statistically, BMRs emerging at higher latitudes tend to exhibit larger tilt angles, popularly known as Joy's law \citep{HE1919, WS89, SG1999, SK2012, Wang2015, MN2016, Poisson20}. The systematic tilt of BMRs is vital for generating the poloidal field for the next cycle and sustaining the solar dynamo \citep{CS15}. Therefore, understanding how BMRs acquire their tilt is essential to understanding the solar cycle and its prediction in this technologically driven society \citep{Petrovay20}.

Simulations complemented by observations support that BMRs originate from magnetically buoyant toroidal flux rising from the deep convection zone \citep{P1955, LG2015}. 
In this scenario, the thin flux tube model \citep{s81, LK97} offers a compelling explanation for the generation of tilt. 
This model assumes that the BMRs are formed through the rise of bundles of magnetic fields of the so-called $\Omega$-shaped loops through the convection zone. As a loop ascends, diverging flows from the apex experience the Coriolis force, imparting a tilt to the BMR \citep{DC1993, Fan93, CM1995}.  
In addition, helical convection, 
 on average, tends to drive tilts to the rising loops in the correct direction \citep{WF2011}.
The thin flux tube model has been central to many simulation studies and has been widely used to investigate various aspects of the solar dynamo and  activity \citep{F2009, Cha20}. 
Beyond the Sun, the model provides insight into the magnetic activity of rapidly rotating stars \citep{isik2018, isik24}, where the Coriolis force dominates the magnetic buoyancy to form polar spots \citep{Schu1992}. 

Despite the extensive implication of the thin flux tube model, its observational support is 
limited
\citep{F2009}.  The thin flux tube model predicts: (i) a reduction of BMR tilt 
with the increase of the toroidal magnetic field as the strong-field flux tubes rise rapidly and the Coriolis force gets less time to tilt and (ii) the increase of tilt with increase of flux due to increasing effects of the drag and Coriolis forces. More explicitly, \citet{Fan93} predicted that the tilt $\gamma \propto \sin \lambda B_0^{-5/4} \Phi^{1/4} $ (where $\lambda$ is the latitude, $B_0$ is the initial magnetic field of the flux tube and $\Phi$ is the flux inside the rising flux tube). While $B_0$ has little or no bearing on the observed photospheric field,  $\Phi$ is related to the measured flux in BMR which has a wide variation in magnitude (more than 3 orders). Hence, we expect a measurable change in the tilt with the magnetic flux. However, simulations considering the effect of convection show that this dependency is not robust \citep{WF2013}. Probably this is why, in observations, no systematic trend is observed; some groups \citep{SK2008, SK2012} find no statistically significant variation in tilt with the BMR flux, while other groups \citep{TL2003, JK2020, SJ2024} find an increase of tilt 
with the flux
followed by a reduction at high flux values.
Thus, we need to check an alternative robust prediction of the thin flux tube model, and that is, Joy's law tilts at the emergence.  The model predicts that BMRs should emerge with a systematic tilt given by Joy's law, as during the journey of flux tubes inside the convection zone itself the Coriolis force induces tilt. However, in contrast to this prediction, previous observations \citep{Howard96, SK2008, SB2020} find that BMRs emerge with nearly zero tilt and acquire the tilt predicted by Joy's law only during the post-emergent phase, leading to the conclusion that 
the tilt is produced in the surface layer rather than below the surface.
Identifying the tilt at emergence remains a challenging problem, as surface turbulence can distort the initial orientation and, as a result, the characteristic Joy’s law signature may not be evident at the early stage of BMR evolution. We track BMRs in the high-resolution magnetogram data over the last two solar cycles, starting from the time when they begin to show signatures in the line-of-sight magnetic field, and find that BMRs display tilt according to Joy's law.

\section{Results}\label{sec:result}

To investigate BMR tilt at the time of emergence, we built on the catalog produced by the Automatic Tracking Algorithm for Bipolar Magnetic Regions \citep[AutoTAB;][]{SJ2023}. This algorithm identifies and tracks BMRs throughout their nearside evolution using line-of-sight (LOS) magnetograms from the Michelson Doppler Imager (MDI) onboard Solar and Heliospheric Observatory (SOHO) and the Helioseismic and Magnetic Imager (HMI) onboard Solar Dynamic Observatory (SDO). The resulting catalog contains 11,987 unique BMRs spanning September 1996 to December 2023, covering solar cycles 23 and 24 in full and part of cycle 25.

\begin{figure}
\centering
\includegraphics[width=0.75\textwidth]{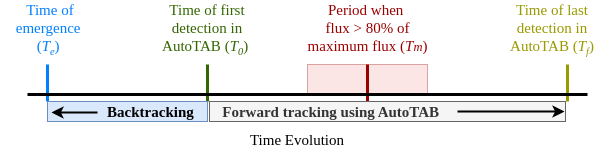}
\caption{Timeline of the BMR evolution, illustrating two tracking phases: the backtracking and  forward tracking. The backtracking phase traces BMRs from $T_0$ (AutoTAB's first detection) to the very initial emergence $T_e$. Forward tracking follows a BMR's growth from $T_0$ through $T_m$ when the unsigned flux reaches its peak to the end of tracking $T_f$ ($T_e < T_0 < T_m < T_f$).}
\label{fig:timeline}
\end{figure}

AutoTAB identifies BMRs in two steps: it first detects bipolar regions with strong flux that satisfy a flux balance condition \citep{SK2012}, and then tracks their evolution across the solar disk using a feature-association technique \citep{JP2021}. However, AutoTAB’s original implementation is optimized for studying the evolutionary properties of BMRs. It initiates tracking only once the flux between the polarities is roughly balanced, making it unfit for identifying the earliest emergence when there is usually a significant flux imbalance.
Hence, we begin with the AutoTAB’s first detection of a BMR defined as time \tO\ and track it backward in time 
to identify its earliest appearance in LOS magnetogram observations, to time \te, the actual emergence time (see Figure~\ref{fig:timeline} for the evolution timeline). 
%
Backtracking of the regions begins from the first AutoTAB detection ($T_0$). We extract region of interest for the intended BMR to be backtracked and measure two quantities, the pixel area (i.e., number of pixels with flux value greater than 100~G) and the total unsigned magnetic flux. We exclude regions first detected beyond $-35\degree$ longitude by AutoTAB from backtracking for minimizing projection effects. We then step backward through earlier magnetograms, sequentially, by differentially rotating the region of interest to capture its earlier evolution. At every step, the pixel area and total unsigned flux is computed and a successful backtracking step requires that the unsigned flux does not decrease by more than 40\% and the pixel area does not decrease by more than 50\% relative to their values at $T_0$. This criterion reflects the physical expectation that an emerging region should grow in flux between its first appearance and $T_0$. Steps showing unsigned flux and pixel area growing in comparison to $T_0$ or the unsigned flux/pixel area drops below 40\%/50\% of $T_0$ (indicating that the region is approaching the noise level) is flagged as unsuccessful. The backtracking is terminated if five such unsuccessful steps occur, or if the region reaches the limb ($-45\degree$ longitude). The last physically consistent step is recorded as the emergence time ($T_e$). Finer details of the backtracking procedure and the choice of thresholds are discussed in \citet{sreedevi25b}.
Despite this, the emergences of some BMRs remain undetected because they either appeared near the east limb, appeared on the Sun's far side, or were already in a decaying phase upon first detection on the nearside. 
Of 11,987 BMRs initially present in the AutoTAB catalog, we could successfully backtrack 3,012 BMRs starting from $T_0$ to $T_e$. Figure~\ref{fig:hist_lt} shows the distributions of the time intervals 
$T_m- T_e$ for BMRs, where $T_e$ is the emergence time timestamp identified by backtracking, and $T_m$ represents the timestamp at matured phase of the BMR at a later time. Of these 3,012 BMRs, we found that about $27\%$ of them do not belong to growing or emerging BMRs for which the unsigned flux does not increase over the time of their tracking.
Therefore, they may not represent the true emerging BMRs whose tilt at emergence is of our interest, and thus to keep our analysis clean, we exclude these ambiguous BMRs from our analysis based on the change in the flux at $T_0$ with respect to the flux at $T_e$. A detailed review of the code along with the study of the extraction of these non-growing/non-emerging BMRs including their properties is presented in a follow-up publication \citep{sreedevi25b}. 

\begin{figure}
	\centering
    \includegraphics[width=\textwidth]{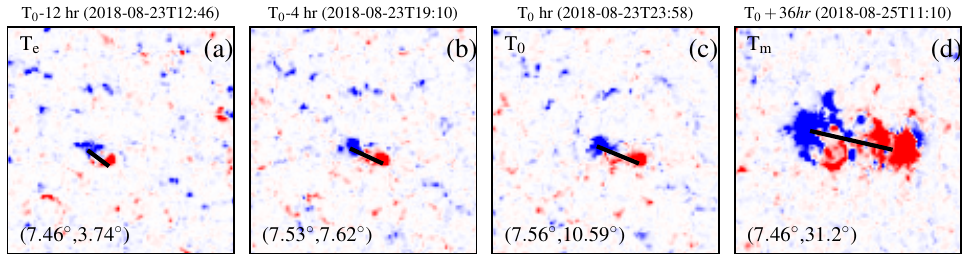}
    \includegraphics[width=\textwidth]{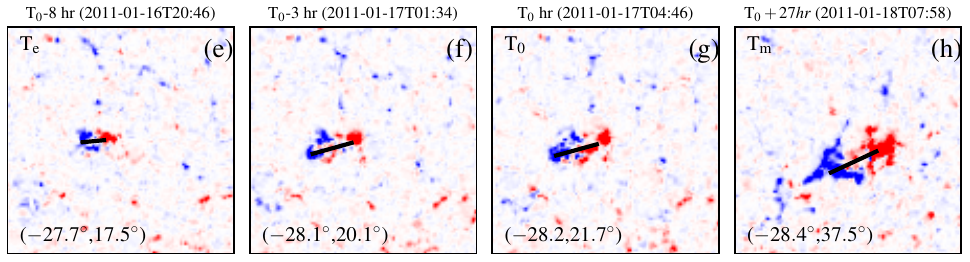}
\caption{
Evolution of two typical BMRs.
(a) and (c), Respectively represent the BMR at the times of initial detection ($T_{e}$) and the starting of our backtracking phase ($T_0$); see Figure~\ref{fig:timeline} for the timeline. (b), The BMR at the middle of the backtracking phase. (d), The same BMR but for comparison at its maximum flux.   
(e--h)
The same as the top row but for a different BMR.
The corresponding (near simultaneous) intensity continuum for these two BMRs are shown in \Fig{fig:IC}. Numbers in brackets on each panel denote the mean latitude and longitude of the region. The magnetic field is saturated at 1.5~kG in all panels.  The line in each panel connects the flux-weighted centroids of BMR's poles. Each box is of size 145~Mm $\times$ 145~Mm. 
The Movie~S1 shows a detailed evolution of (a--d).
}
	\label{fig:bt_eg}
\end{figure}

To show how well our code captures the emerging phases of BMRs, we show the time evolution of two BMRs in Figure~\ref{fig:bt_eg}. As evident in this figure, the code detects the very early emergence phase of these BMRs when they show faint signatures in the magnetogram and almost no signature in Intensity Continuum (Appendix \Fig{fig:IC}) and effectively tracks them as long as they are on the near side of the solar disk. From the time of initial emergence at \te\ to AutoTAB's first detection \tO\, we find that both the measured magnetic flux and footpoint separation generally increase in a steady and coherent manner. This behavior is characteristic of emerging or growing BMRs
{\bf\blue{ \citep{svanda25} }}
and further confirms our robustness in the identification of BMR's early emergence phase.

\begin{figure}
\centering
\includegraphics[width=0.6\textwidth]{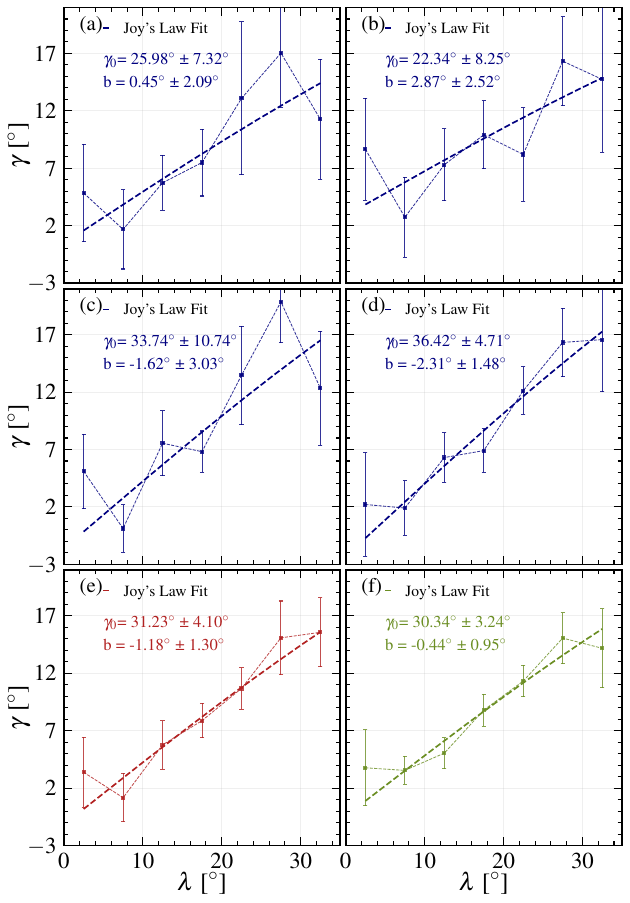}
\caption{
Joy's law (latitude variation of the tilt angle) at different stages of BMR's evolution. (a), At the initial emergence $T_e$.  (b--e) Respectively represent tilt and Joy's law fit at approximately 25\%, 50\%, 75\%, and 100\% (i.e., at $T_0$) of 
the backtracked phase. (Hence, the panel (c) presents the middle of the backtracking phase.)
(f), At the matured phase ($T_{m}$). We note that (a), (e), and (f) include the whole usable BMRs (1876), while the temporal bins in (b), (c), and (d) are chosen in such a way that they accommodate an equal number of BMRs ($\sim 1500$). The error bars are computed from the Gaussian fitting of the tilt data in each latitude bin. 
}
\label{fig:jl} 
\end{figure}

Next, we measure the tilt angles relative to the east-west equatorial line using the flux-weighted centroids, of the two polarities ensuring that the values remain within $\pm90^\circ$. Here we emphasise that, pixels whose magnetic field values exceeds 100~G are only considered for determining the centroids. We adopt the definition for the tilt angle $\gamma$ defined as  
$\tan \gamma = \Delta \lambda / \left((\Delta \phi)\cos \bar\lambda\right)$
\citep{WS89}, wherein $\Delta \lambda$ and $\Delta \phi$ represents the latitudinal and longitudinal separation respectively and $\bar\lambda$ represent the mean latitude of the centroids. This definition ensures
BMRs located in the northern hemisphere and exhibiting Hale–Joy orientations are assigned positive tilt values. In other words, orientations that facilitate/hinder the polar field reversal are considered positive/negative \citep{KM18}. To standardize comparisons, we assume hemispherical symmetry by reversing the sign of tilts for all BMRs emerging in the southern hemisphere.

With this standardized definition, we move on to examine the latitude dependence of the tilt angle at $T_e$ from the usable data. For this, we group the tilt data into $5\degree$ latitude bins and calculate the mean and uncertainty by fitting the tilt in each bin using a Gaussian function. 
These values as a function of the latitude bin centers are presented in Figure~\ref{fig:jl}(a). After fitting the data to the standard Joy's law equation, $\gamma = \gamma_{0}\sin\lambda + b$, we obtain a Joy's law amplitude of $\gamma_0 = 25.98\degree \pm 7.32\degree$. We get comparable results if we compute the median tilt instead of the Gaussian mean tilt in each latitude bin. As seen in Figure~\ref{fig:jl}a, although there is a considerable amount of spread in each latitude bin of the tilt, Joy's law dependence is statistically robust and significant. 

We recall that in the present work, we have tracked the BMRs detected by our previous AutoTAB code back in time to capture their earliest possible signatures in the LOS magnetograms, because previously, due to 
imposed flux balance criterion of $(|\Phi_{+}| - |\Phi_{-}|)/(|\Phi_{+}| + |\Phi_{-}|) < 0.4$, where $\Phi_\pm$ denotes the total unsigned flux of each polarity, AutoTAB could not detect the initial phase of the majority of BMRs.
However, we found that for 307 BMRs, previous AutoTAB could indeed capture the initial emergence, i.e., for those 307 BMRs, $T_e = T_0$. We excluded these BMRs in the present study of tilt properties. Interestingly, even if we include these 307 BMRs, the tilt properties, particularly, Joy's law at $T_e$, remain consistent as shown in \Fig{fig:jlall}.

We observe that as BMRs grow with progressive flux emergence, they become more stable and Joy's law tends to become more and more prominent. In Figure~\ref{fig:jl}, we observe a progressive strengthening of Joy’s law over time, accompanied by a decrease in scatter around the mean tilt in each latitude bin.  At approximately 75\% of the backtracking period, corresponding to an average BMR age of 0.78 days, the scatter decreased considerably (Figure~\ref{fig:jl}d). Moving forward in time, at $T_0$, i.e., at the starting of our bactracking phase, the Joy's law trend is stable and is consistent with what we found in our previous work using the AutoTAB catalog \citep[Figure 5 of ][]{SJ2024}.  Finally, the measured Joy's law at the time of maximum flux as presented in panel (f) is in agreement with our previous studies \citep{SJ2024}, and the studies that do not track BMRs (and thus measure tilt at any phase of a BMR and the same BMR is considered repeatedly) \citep[e.g.,][]{SK2012, MN2016, JK2020}. 

\begin{figure}
       \centering
       \includegraphics[width=1.0\textwidth]{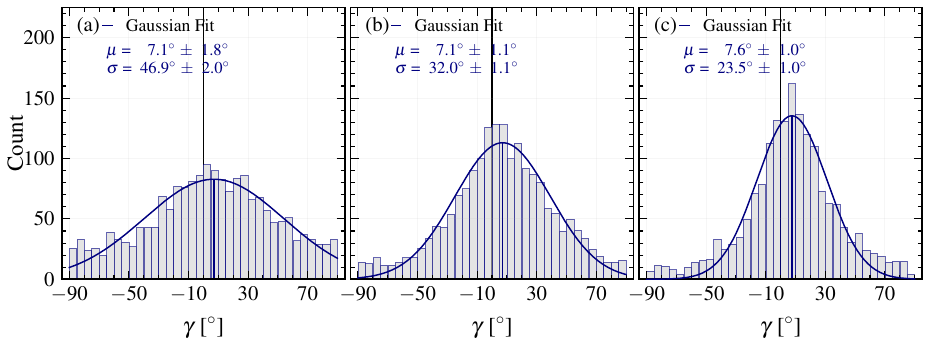}
       \caption{Tilt distribution at different times. (a), (b) and (c) Respectively represent the distributions of the BMR tilt angles at $T_e$, $T_0$, and $T_m$. The parameters of the Gaussian fit, $\mu$ and $\sigma$ are printed on each panel.}
	\label{sfig:tiltdist}
\end{figure}

The reduction of scatter around the mean Joy's law becomes even more apparent when we examine the overall tilt distribution from all latitudes at different times. In Figure~\ref{sfig:tiltdist}, we find that the mean of the distribution remains relatively stable from \te\ to \tO\ to \tm. It is the spread that narrows significantly over time. This reduction of scatter over the BMR time evolution was previously reported by \citet{SK2008, SB2020}; however, they could not detect any Joy's law trend at the time of emergence. 
It is this huge scatter at the time of emergence that tends to hide the Joy's law signature. Previous studies with limited and low-resolution data could possibly have resulted in no statistical signature of Joy's law. 
To demonstrate it more specifically, in Appendix (\Fig{fig:iteration}), we show the distribution of mean tilt angles of 100 randomly selected BMRs with flux $10^{22}$~Mx. We observe that there are cases where the mean goes to nearly zero or even negative, suggesting that a cause of zero tilt at the time of emergence found in the analysis of \cite{SB2020} using 153 BMRs could be a poor statistic.

\begin{figure}
\centering
\includegraphics[width=0.7\textwidth]{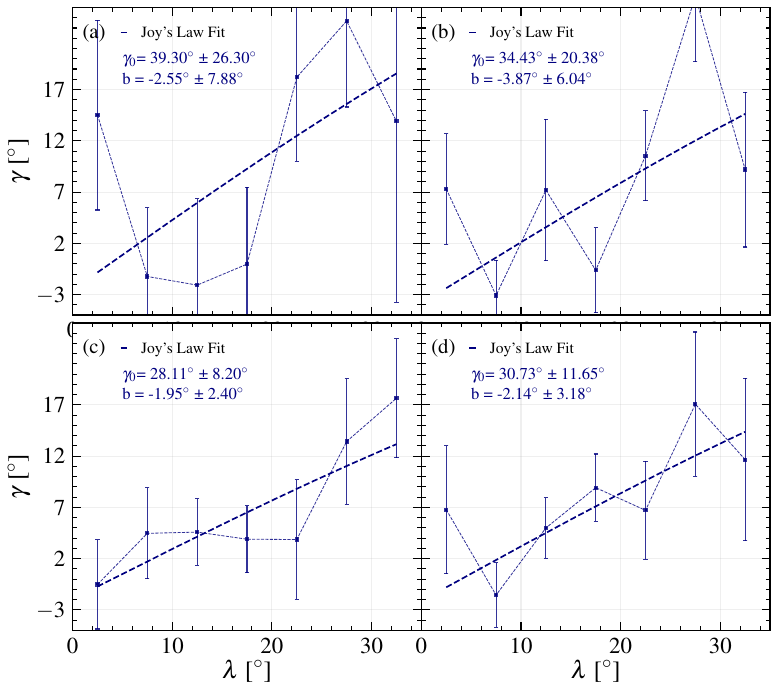}
\caption{       
Evolution of Joy's law with the flux emergence during BMR's early phase.  Joy's law computed when the BMR flux falls between (a) 0--5\%, (b) 0--10\%, (c) 5--15\%, and (d) 20--30\% of the maximum flux. No of BMRs belonging to (a) to (d) are 436, 729, 809, and 862.  It is essential to remember that due to the unequal growth rates of the BMRs, we capture the different phases of BMRs at the fixed flux range.
}
\label{sfig:jlflux}
\end{figure}

Finally, in terms of the flux, a similar strengthening of Joy's law over the BMR evolution is seen. Even the BMRs at $T_e$ having the flux only $5\%$ of the maximum value show a signature of Joy's law; Figure~\ref{sfig:jlflux}(a). We find that by the time 10\% flux emerges on the surface, BMRs display a strong Joy's law with reduced scatter around it; Figure~\ref{sfig:jlflux}(b). With the increase of more flux emergence, the scatter decreases even further and the Joy's law trend remains more or less consistent with the one obtained at the time of maximum flux (Figure~\ref{sfig:jlflux}c-d) 
This is trend is in agreement with \cite{will24}, who found strengthening of Joy's law from the time when $20\%$ of the flux has emerged to the time of maximum flux.

\section{Conclusions and Discussion}\label{sec:conclusion}
By carefully backtracking the BMRs detected in the LOS magnetograms of more than last two solar cycles using AutoTAB \citep{SJ2023} till the time of their emergence on the solar surface, we for the first time robustly present Joy's law of the tilt angle at the time of emergence $T_e$. The scatter around Joy's law or the spread of the tilt distribution is rather high at the time of BMR emergence. However, this scatter reduces in time over the post-emergence phase, in agreement with previous reports \citep{SK2008, SB2020} and by the time when the BMRs acquire a certain amount of balanced flux between two opposite poles which corresponds to the starting of the backtracking phase ($T_0$), the scatter is reduced significantly and Joy's law becomes extremely tight.  

This observed Joy's law at $T_e$ and the subsequent reduction of scatter around Joy's law during the emergence phase of BMRs (from $T_e$ to $T_0$) closely align with the theoretical prediction based on the thin flux tube model \citep{LC2002, WF2011, WF2013}. The theory showed that BMRs acquire systematic tilts through Coriolis force during the journey of BMR-forming flux tubes through the convection zone \citep{DC1993}. However, the orientations of the flux tubes are significantly disturbed during their travel through the highly turbulent near-surface layer, causing a departure from the expected Joy's law \citep{LF1996}. 
Turbulence perturbations relax on shorter time scales than the time scale of perturbations from the Coriolis force, 
allowing turbulence effects to dissipate rapidly.  Hence, we expect the scatter around Joy's law to be maximum at the first emergence on the photosphere and to reduce after a relaxation. In our data, we are getting exactly similar behavior; a significant amount of scatter is reduced in about a day (the maximum and average durations of our backtracking phase are 4.13 and 1.25 days; Figure \ref{fig:jl}).

Observations show that the convective flows control the dynamics of rising flux tubes in the 
near-surface layer 
of the Sun \citep{Brich16}. 
However, the thin flux tube model does not include the interaction of the flux tubes with the convection and thus not every detailed prediction of this theory is expected to be held in the observation. For example, when convection is included in thin flux tube rise simulations, in addition to the Coriolis force, helical convection (negative/positive in the northern/southern hemisphere)  in the solar convection zone was found to increase tilt in accordance with Joy’s law \citep{WF2011, WF2013}. 
Another consequence of convection is that it diminishes the flux dependence of the tilt as predicted by the thin flux tube model without convection \citep{FF1995}. Cartesian simulation of the flux tube rise 
in the near-surface convection zone
\citep{hannah25} shows that the Coriolis force can induce a tilt consistent with Joy's law by acting on the horizontal flows.  Thus, more realistic three-dimensional simulations of the flux tube rise in the deep convection zone \citep[such as by][]{JB09, HI20} and the full MHD spherical-shell convection \citep[e.g.,][]{HRY16, Hotta19}, are needed for the detailed mechanism of the flux tube rise in the solar convection zone. However, the observed Joy's law tilt at the first emergence suggests that a large portion of the active region tilts are already developed before they imprint signature on the solar surface and are not caused by surface flows. The possible cause of this tilt is the Coriolis force and the helical convection.  

\begin{acknowledgments}
We thank Arnab Rai Choudhuri, Robert Cameron, Lisa Upton, Kuldeep Verma,  Prasun Dutta and anonymous referees for providing valuable comments and suggestions on the results, which helped to improve the quality of the work. B.B.K. acknowledges the financial support from the Indian Space Research Organisation (project no. ISRO/RES/RAC-S/IITBHU/2024-25) and the Anusandhan National Research Foundation (ANRF) through the MATRIC program (file no. MTR/2023/000670).
\end{acknowledgments}

\begin{contribution}
B.B.K. gave the initial idea and drove this project. A.S., under the guidance of B.B.K., B.K.J., and D.B. analyzed magnetogram data, developed algorithms for BMR back tracking, and prepared the figures. R.G. analyzed Intensity Continuum data and helped in preparing some figures.    B.B.K. and A.S. led the manuscript writing while B.K.J. commented on the text. All authors contributed to the discussion and presentation of the results. 
\end{contribution}

\bibliography{sample701}{}
\bibliographystyle{aasjournalv7}

\section*{Appendix}
\begin{figure}[h!]
\centering
\includegraphics[width=1.0\textwidth]{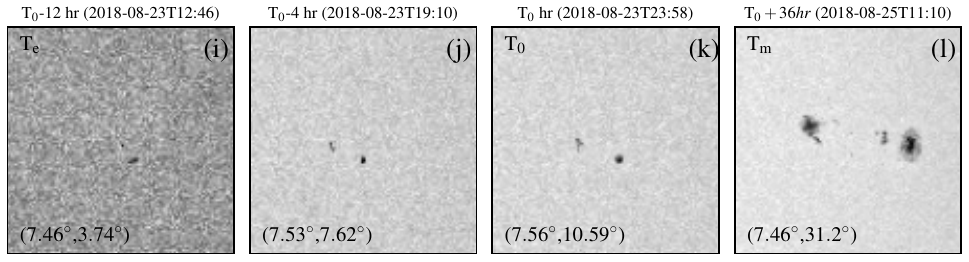}
\includegraphics[width=1.0\textwidth]{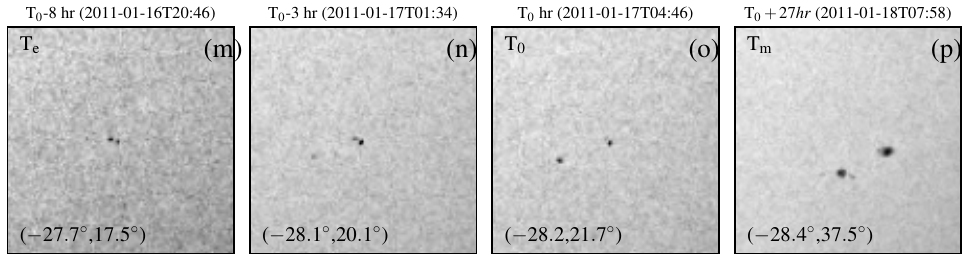}
\caption{
(i--l), and  (m--p) The same BMRs that are shown in \Fig{fig:bt_eg} are now searched in the corresponding (near simultaneous) intensity continuum.
}
\label{fig:IC}
\end{figure}

\begin{figure}
    \includegraphics[width=0.5\textwidth]{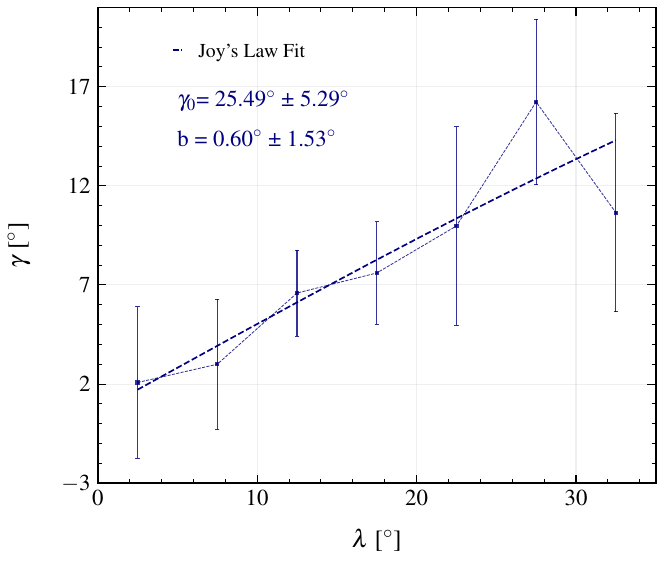}
    \caption{
The same as \Fig{fig:jl}(a) but including those additional 307 BMRs for which $T_0$ was the same as $T_e$.     
}
\label{fig:jlall}
\end{figure}

\begin{figure*}
    \includegraphics[width=\textwidth]{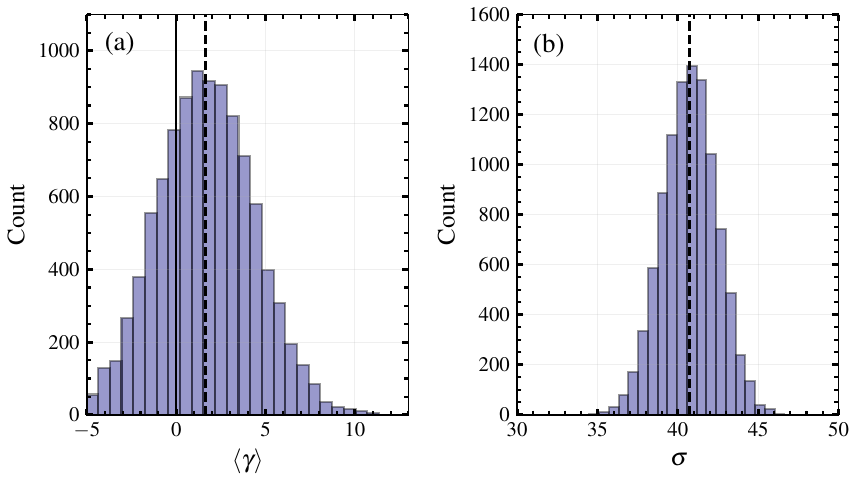}
    \caption{
    The distributions of 10000 the (a) means and (b) standard deviations of the tilts at $T_e$. Each mean/standard deviation is computed from randomly picked 100 BMRs having matured flux $\ge  10^{22}$~Mx \citep[to be close to the emerging active regions used in][]{SB2020}. }
\label{fig:iteration}
\end{figure*}

\begin{figure}
    \includegraphics[width=0.5\textwidth]{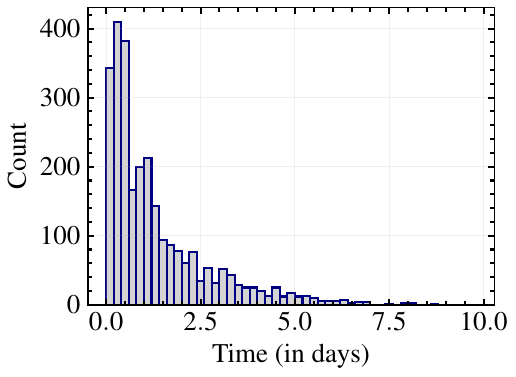}
    \caption{Distribution of the time intervals $T_m-T_e$ for BMRs, where $T_e$ is the emergence time identified by backtracking and $T_m$ is the later time when the BMR reaches its maximum flux.}
\label{fig:hist_lt}
\end{figure}

This section presents additional information to strengthen the results presented in the Letter. In particular, we show the time evolution of the same BMRs that are presented in \Fig{fig:bt_eg} but in the intensity continuum where we can observe the sunspots. As shown in \Fig{fig:IC}, we find that our detection code clearly captures the early emergence phase of the sunspots, in particular, a few hours before they show imprints on the surface in the continuum.  
Next, we show the sensitivity of Joy's law when including the additional 307 BMRs for which $T_e = T_0$. As seen in \Fig{fig:jlall}, Joy's law trend remains almost the same as found for the BMRs excluding these BMRs in \Fig{fig:jl}(a). Finally, in \Fig{fig:iteration}, we present the distributions of the mean and standard deviation of the 100 tilt data at $T_e$ randomly selected from the entire data set used in our study. This shows that the tilt properties at $T_e$ vary quite a bit with the number of BMRs used. If we have fewer BMRs (here 100), then there is a high chance that the mean can be even near zero, which is not the case for the Sun.   

\end{document}